\documentclass[a4paper]{spie}  

\usepackage{amsmath,amsfonts,amssymb}
\usepackage{graphicx}
\usepackage[colorlinks=true, allcolors=blue]{hyperref}
\usepackage{xcolor} 

\title{Microchannel plate detector development for ultraviolet astronomy}

\author[a]{S.~Diebold}
\author[a]{J.~Barnstedt}
\author[a]{J.~Bayer}
\author[a]{L.~Conti}
\author[b]{H.R.~Elsener}
\author[a]{C.~Kalkuhl}
\author[c]{D.~Rau}
\author[c]{D.~Schaadt}
\author[a]{T.~Schanz}
\author[a]{B.~Stelzer}
\author[a]{K.~Werner}

\affil[a]{Institut für Astronomie und Astrophysik, Eberhard Karls Universität Tübingen, Sand 1, 72076 Tübingen, Germany}
\affil[b]{Empa, Swiss Federal Laboratories for Materials Science and Technology, Ueberlandstrasse~129, 8600 Dübendorf, Switzerland}
\affil[c]{Institut für Energieforschung und Physikalische Technologien, Technische Universität Clausthal, Leibnizstraße 4, 38678 Clausthal-Zellerfeld, Germany}

\authorinfo{Corresponding author: Sebastian Diebold\\E-mail: diebold@astro.uni-tuebingen.de, Telephone: +49 7071 29 78604}

\begin{document}
\maketitle

\begin{abstract}
Observational data from the UV wavelength range is crucial to solve several astrophysical puzzles. Consequently, there are a number of upcoming UV missions from CubeSats up to the future flagship Habitable Worlds Observatory. Besides novel instrument concepts and improved coatings, advanced detectors are key for the success of these missions. The microchannel plate (MCP) detector technology offers a unique asset in the UV: the combination of single-photon counting and visible-blindness.

The UV hardware group at the Institut für Astronomie und Astrophysik Tübingen (IAAT) develops a versatile MCP detector system that addresses the complete UV band. While a sealed-tube design is limited to wavelengths above 118\,nm, an open-face variant also covers the whole far- and extreme-UV. Both use the same readout: a coplanar cross-strip anode and FPGA-based electronics. In this contribution, we report on the status of the detector development, present the latest characterization results, and give an outlook on the mission prospects.
\end{abstract}

\keywords{UV instrumentation, UV detector system, microchannel plates, MCP detector, AlGaN photocathode, cross-strip anode, centroiding algorithm}

\section{INTRODUCTION}
\label{sec:intro} 

A detector system based on microchannel plates (MCPs) allows to combine single-photon counting capability with visible-blindness and is, thus, perfectly suited for ultraviolet (UV) wavelengths with usually low count-rates and possible straylight 
issues. The bandgap of the chosen photocathode defines the long-wavelength sensitivity cut-off and is for UV instruments usually sufficiently large that thermal excitiation is negligible and, hence, cooling can be omitted.

MCP detectors have a long track record of successful applications in UV astronomy
\textcolor{violet}{. Despite} 
the considerable advancements over the past 30 years in making silicon technology applicable for the UV, MCP-based detectors are still considered for several future UV missions, from CubeSats up to a flagship like the Habitable Worlds Observatory (HWO)\cite{Curtis2025}. A detailed explanation of their functional principle as well as a brief compilation of the history of MCP-based detectors in astronomy is given in Diebold (2022)\cite{Diebold2022}. Already since the 1980s, MCP detectors are developed at the Institut für Astronomie und Astrophysik of the University of Tübingen (IAAT). These efforts continue at IAAT, and led to a state-of-the-art position-sensitive detector with cross-strip anode and FPGA-based readout electronics.

This publication highlights the current fields of further improvement of the IAAT detector system as well as the mission prospects for its application. In the following Section~\ref{sec:mission}, two recently proposed missions are outlined that are foreseen to be equipped with IAAT MCP detectors. In Section~\ref{sec:status}, our twofold detector design is briefly introduced, before our latest results on growing an (Al)GaN photocathode directly on MCPs are summarized, and the implementation of a centroiding algorithm for our cross-strip anode in the FPGA of our readout electronics is sketched.

\section{MISSION PROSPECTS}
\label{sec:mission}
Two missions that specifically drove the development of our detector system have been thoroughly described before and are only briefly mentioned here: the \textit{ESBO} design study (DS) for a reusable near-UV balloon instrument \cite{Pahler2020,Bougueroua2022,Vernoosfaderani2022} and the \textit{TINI} project for a far-UV payload on a commercial or scientific space platform \cite{Diebold2022a}. Both projects are on hold for the moment since for \textit{ESBO DS} no funding is available for a commissioning flight and \textit{TINI} was not finally adopted by the Indian space agency ISRO. Efforts are ongoing to realize these missions through institutional, public, or private funding.

However, recently two European missions -- \textit{SIRIUS} and \textit{HYADES} -- that include the IAAT MCP detector system advanced to the second stage of the selection process for ESA's next fast mission (F3). Therefore, a brief overview of \textit{SIRIUS} and \textit{HYADES} is given in the following.

\subsection{SIRIUS}
\label{ssec:sirius}
\textit{SIRIUS} \textit{(Stellar and Ism Research via In-orbit Ultraviolet Spectroscopy)} targets high-resolution spectroscopy of stars and the local interstellar medium in the extreme-UV (EUV) wavelength range 17 -- 26\,nm. It will be the first mission dedicated to extra-solar EUV astronomy since \textit{EUVE}\cite{Bowyer1994} but with an effective area up to one magnitude larger and, thus, a significant increase in sensitivity.

The mission is proposed by an international team under the lead of Leicester University (PI: M.\@ Barstow) and with British industry partners. \textit{SIRIUS}'s highly efficient instrument with only one single focusing and dispersive optical element is a further development from the successful \textit{J-PEX} sounding rocket program\cite{Barstow2014}. The project's excellent science objectives and technical maturity were highlighted by the selection as backup mission for ESA's second fast mission F2. Afterwards, it was funded by UKSA for a phase-0 study that comprised the development of an expandable optical bench that is compatible with commercial satellite buses. The mission was then proposed in ESA's exploratory mini-F call in 2025, but was moved to the second stage of the current fast mission call (ESA F3). The F3 selection is announced for November 2026.

More details on \textit{SIRIUS} and the preceding \textit{J-PEX} flights can be found in Barstow et al.\@ (2014)\cite{Barstow2014} and the references therein. The latest updates for the \textit{SIRIUS} mission and a summary of the results from the recent UKSA phase-0 study can be found in Barstow et al.\@ (2026), included in this proceedings volume\cite{Barstow2026}.

\subsection{HYADES}
\label{ssec:hyades}

\textit{HYADES} \textit{(HYdrogen And DEuterium Surveyor)} is a proposed space mission to investigate the origin of Earth's water and to search for previously unknown sources of water on minor bodies in the Solar System and beyond. The objective is to quantify the deuterium-to-hydrogen (D/H) ratio in about 100 comets and 50 comet-like asteroids in the main asteroid belt. Furthermore, \textit{HYADES} will constrain the water-ice sublimation and the D/H ratio for every interstellar object discovered during the mission lifetime, and it will perform a mapping of the water-ice sublimation across the main asteroid belt.

The chosen tracer for D/H is hydroxyl (OD/OH) that is formed via photodissociation of sublimated water (DOH, H$_2$O). The OD/OH line is bright in the near-UV at 308\,nm, and due to its lower abundance, the OD fraction drives the sensitivity requirement. \textit{HYADES} features a 50-cm telescope equipped with narrow-band filters on a filter wheel. Two observing modes are realized with a high-throughput imager and a high-resolution long-slit spectrograph. Both instruments apply the IAAT sealed-tube MCP detector system with a photocathode optimized for 308\,nm. Low count-rate observations will be performed in photon-counting mode, while for high count-rates images can be directly integrated in the readout electronics.

\textit{HYADES} is proposed by an international team under the lead of Jagellonian Universiy Krakow (PI: M. Drahus) and with strong contributions from the Polish company CBK-PAN. The development of the mission design was funded by the European Research Council under an ERC Consolidator Grant. Similarly to \textit{SIRIUS}, \textit{HYADES} was first proposed in ESA's recent exploratory call for mini-F missions, and subsequently moved by ESA to the second stage of the selection process for the third fast mission (F3).

An overview of the mission as well as latest news can be found on the \textit{HYADES} website.\footnote{\url{https://www.hyades.oa.uj.edu.pl/}}

\section{DETECTOR DESIGN AND STATUS}
\label{sec:status}

The IAAT MCP detector system consists of three units: \textit{1)} the light sensitive detector head with photocathode, MCP stack, coplanar cross-strip anode, and frontend pre-amplifier electronics; \textit{2)} the FPGA-based readout electronics including the analog-to-digital converters (ADCs) and the external interfaces; and \textit{3)} the high-voltage power supply. Depending on the application and the wavelength range, two options for the detector head are available: \textit{a)} a sealed-tube design with an MgF$_2$ or sapphire window and \textit{b)} an open-face design with a mechanical shutter system. While the short-wavelength cut-off for MgF$_2$ is at about 118\,nm and thus includes the Lyman-alpha line, sapphire becomes opaque already at about 160\,nm. Nevertheless, sapphire brings the advantages of simpler fabrication and handling and is a commonly used substrate for GaN. Furthermore, the sealed design simplifies detector handling and operation, minimizes operational risks such as degradation of the performance by exposition to air, and mitigates a possible failure of the mechanical shutter system.

\subsection{Photocathode development}
\label{ssec:pc}
In order to meet the sensitivity requirements of future FUV and EUV missions like \textit{SIRIUS}, open-face MCP detectors require solar-blind photocathodes and high quantum efficiency (QE). While KBr and CsI are standard for the FUV, ultra-wide bandgap III-nitride semiconductors, specifically Al$_{x}$Ga$_{1-x}$N, offer a significant advantage. By tuning the aluminum fraction $x$, the bandgap can be engineered from about 3.3\,eV (pure GaN) to 6.0\,eV (pure AlN) \cite{Muth1999}, allowing for a customizable long-wavelength cut-off that inherently suppresses the visible background without the need for external filters. Furthermore, optimized p-doped Al$_{x}$Ga$_{1-x}$N combined with a thin cesium-oxide (Cs/O) surface activation layer can achieve negative electron affinity (NEA), facilitating the escape of photoelectrons into the vacuum, leading to high QE. Due to the ultra-wide bandgap of Al$_{x}$Ga$_{1-x}$N, an extremely low dark count rate can be achieved \cite{Cai2021}.

One challenge in developing high-QE Al$_{x}$Ga$_{1-x}$N photocathodes for open-face applications is the lattice mismatch between the substrate and the active film, which introduces stress and defect-driven recombination sites that trap charge carriers. In a recent study, we investigated the molecular beam epitaxy (MBE) growth of Al$_{x}$Ga$_{1-x}$N films ($0 \le x \le 1$) on cubic MgO (1\,0\,0) substrates. Unlike standard sapphire substrates that promote the thermodynamically stable but polar hexagonal (wurtzite) structure, the MgO substrate successfully stabilizes the metastable cubic phases of the III-nitrides. The nonpolar nature of these cubic phases is desirable, as it prevents polarization-induced internal electric fields, thereby increasing carrier lifetime \cite{Diebold2025}.

\begin{figure} [h]
   \begin{center}
        \includegraphics[width=.55\textwidth]{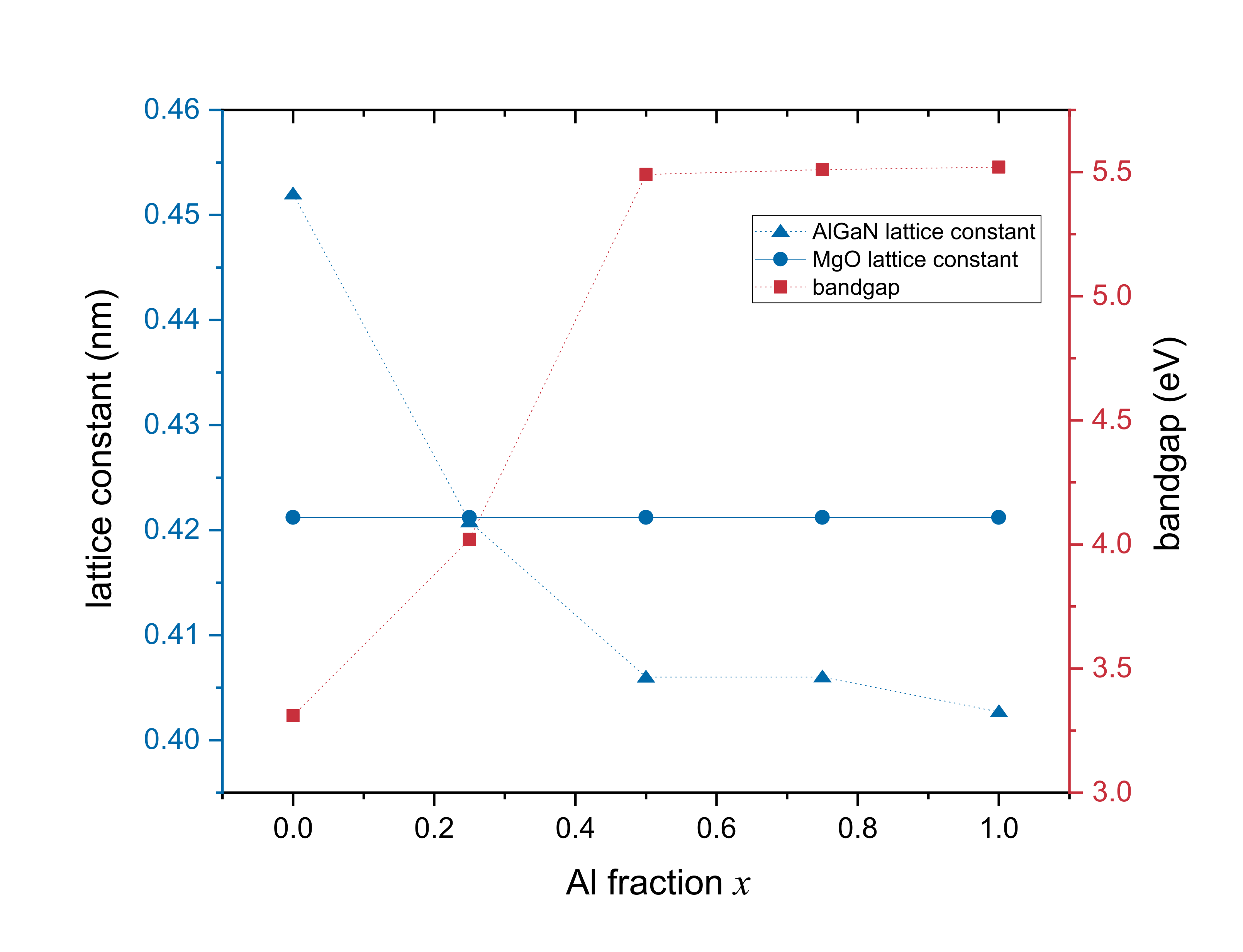}
    \end{center}
    \caption{Lattice constants of the MgO substrate and Al$_{x}$Ga$_{1-x}$N films alongside their measured bandgap for different Al fractions $x$. The lowest lattice mismatch is achieved at $x=0.25$. For lower values of $x$, a strong dependence of the bandgap on the Al fraction is visible, establishing the baseline for precise cut-off tuning in visible-blind UV detectors. Figure adapted from Diebold et al.\@ (2025)\cite{Diebold2025}.}
    \label{fig:lattice_bandgap}
\end{figure}

High-resolution X-ray diffraction (HRXRD) and reciprocal space maps confirmed the absence of hexagonal phases in the compound films. Most notably, we observed the stabilization of the rare cubic rocksalt (gamma) phase at high aluminum concentrations ($>82\,\%$). The structural and optical properties were correlated using UV-Vis-NIR spectroscopy. As illustrated in Figure~\ref{fig:lattice_bandgap}, the measured bandgap closely follows the expected trend for visible- and solar-blindness, saturating near 5.5\,eV for high Al fractions. The optimal structural alignment was found at an initial aluminum fraction of $x=0.25$, yielding an exceptionally low lattice mismatch of approximately 0.1\,\% to the MgO substrate.

\begin{figure} [h]
   \begin{center}
        \includegraphics[width=.95\textwidth]{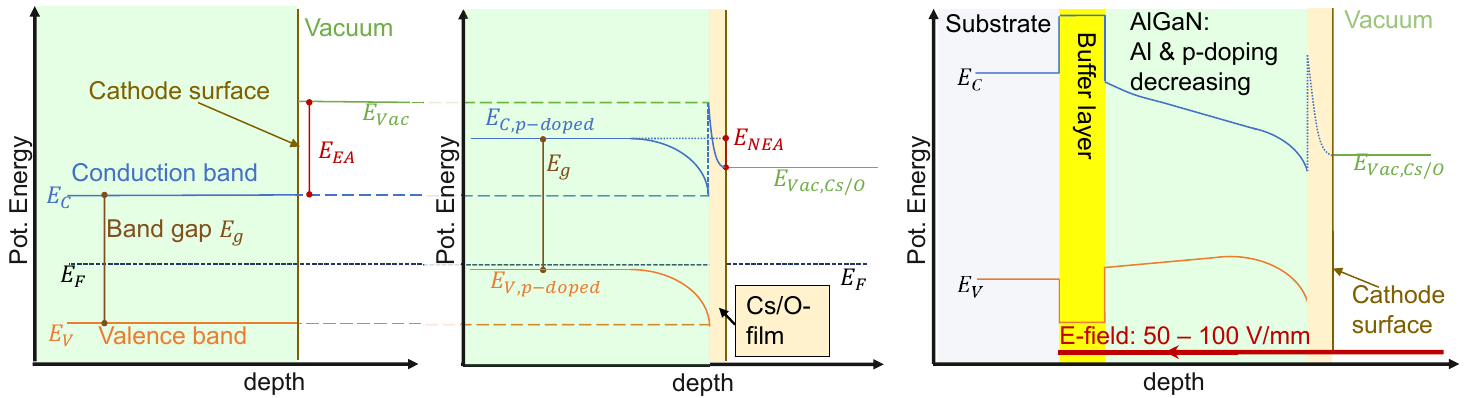}
    \end{center}
    \caption{Effect of p-doping, Cs/O activation, and Al concentration on the band structure of Al$_{x}$Ga$_{1-x}$N photocathodes. Left: undoped. Center: p-doped and Cs/O-activated. Right: gradient p-doping and decreasing Al concentration towards the surface of the NEA photocathode. Figure adapted from Conti (2021)\cite{Contiphd21}.}
    \label{fig:Bandschematics_AlGaN}
\end{figure}

Building upon these material science results, our ongoing detector development focuses on transferring the epitaxial growth process to functional MCPs. The ultimate open-face detector configuration requires the photocathode to be operated in opaque mode, where the Al$_{x}$Ga$_{1-x}$N film is directly deposited onto the MCP surface. As MCPs, we use borosilicate glass capillary arrays (GCAs), which are functionalized via atomic layer deposition (ALD). In configurations utilizing a buried electrode, the secondary electron emissive layer of the ALD-MCP also forms its topmost surface. Because these thin films are inherently amorphous, techniques to deposit crystalline buffer layers onto these films are required. Developing these films to serve as suitable epitaxial templates remains an active field of research \cite{IBAD-MgO}. Future work also encompasses optimizing the p-doping gradients to enhance electron transport towards the surface. As Mg-doping must be optimized for each alloy of GaN and AlN, the gradient is in reality less smooth than sketched in Figure~\ref{fig:Bandschematics_AlGaN}. Furthermore, robust Cs/O activation parameters in ultra-high vacuum (UHV) must be established to achieve NEA and, simultaneously, long-term stability for the fully integrated MCP detector system.

While \textit{SIRIUS} drives the development of open-face detectors, missions targeting the near-UV, such as \textit{HYADES}, require high sensitivity at approximately 300\,nm. For this application, an (Al)GaN photocathode can be deposited directly, for example, onto a sapphire substrate, which simultaneously serves as the sealing window for the closed-tube detector design. The bandgap remains sufficiently wide to maintain exceptionally low dark count rates at room temperature, completely eliminating the need for active cooling systems.

\subsection{Centroiding algorithm}
\label{ssec:centr}

The IAAT detector system employs a LTCC (low-temperature cofired ceramics) cross-strip anode consisting of 64 strips in both coordinate directions over an active area of $39\times39\,\mathrm{mm^2}$. In this coplanar design, the strips encoding the $y$-coordinate are realized as continuous electrodes, whereas the $x$-coordinate is measured using interconnected rectangular pads. This geometry reduces crosstalk and manufacturing complexity while maintaining the excellent spatial resolution characteristic of cross-strip anodes at comparatively low MCP gain.

In order to allow an implementation in our FPGA-based readout electronics, we recently developed a non-iterative centroiding algorithm for our cross-strip anode. The motivation was to combine high spatial accuracy with deterministic execution time.

The centroiding algorithm assumes that the charge cloud generated by the MCP stack can be approximated by a Gaussian distribution. Instead of determining the centroid position using an iterative Gaussian fit, the Gaussian function is transformed by applying the natural logarithm, resulting in a quadratic polynomial\cite{Caruana1986}. Following the weighted formulation proposed by Guo (2012)\cite{Guo2012}, the polynomial coefficients can be determined from a system of linear equations, from which the centroid position is calculated directly. Since only fixed arithmetic operations are required, the algorithm can be implemented entirely using integer arithmetic with deterministic runtime.

In order to increase robustness, all strips exceeding a predefined threshold contribute to the centroid calculation. In contrast to earlier three-point approaches, this allows a larger fraction of the measured charge distribution to be used without significantly increasing the computational effort. The implementation was further optimized for FPGA hardware by replacing logarithmic calculations with lookup tables and by processing the two detector coordinates in parallel.

\begin{figure}[t]
    \centering
    \includegraphics[width=\linewidth]{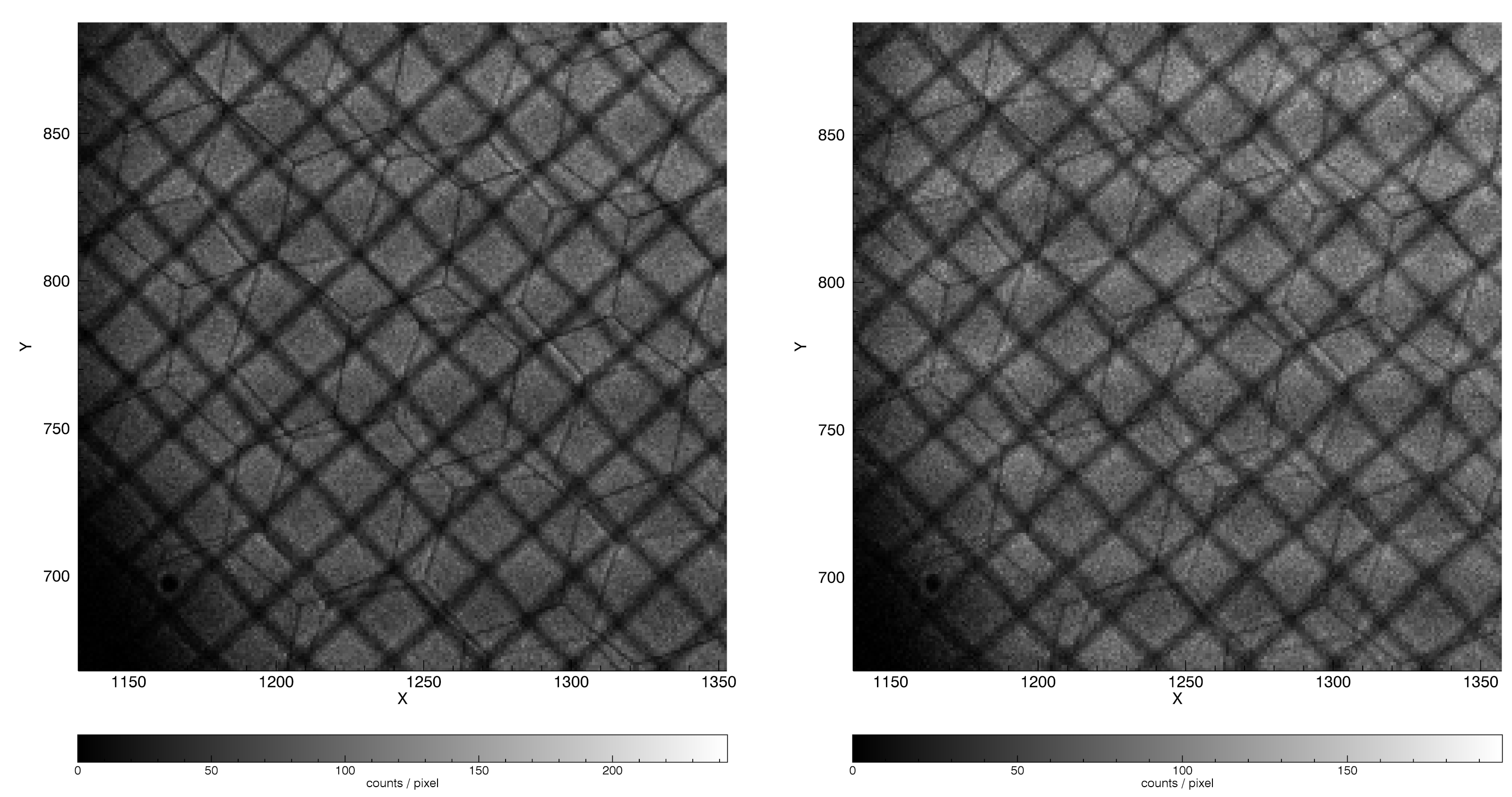}
    \caption{Comparison between Gaussian fitting (left) and the proposed
    non-iterative centroiding algorithm (right). The non-iterative approach
    preserves the fine detector structures while considerably reducing the
    computational complexity. Figure adapted from Diebold et al.\@ (2025)\cite{Diebold2025}.}
    \label{fig:gauss_comp}
\end{figure}

The proposed algorithm was first evaluated in software using recorded detector events and compared to conventional Gaussian fitting. Figure~\ref{fig:gauss_comp} illustrates that both methods reconstruct the detector image with similar quality. Fine detector structures, including the hexagonal MCP substructure, remain clearly visible using the non-iterative approach, indicating only a minor reduction in spatial resolution compared to the computationally more expensive Gaussian fit. The average deviation between both centroiding methods was found to be well below one reconstructed pixel.

To verify the hardware implementation, the same data set was processed by an FPGA implementation of the algorithm. The software implementation reproduced the Gaussian fit reference with almost identical accuracy, whereas the first FPGA realization exhibited slightly larger deviations due to the limited numerical precision available in the processing pipeline. Nevertheless, this preliminary implementation demonstrated the feasibility of real-time centroiding and already reached event processing rates of approximately $2.8\times10^4$ events per second.

\begin{figure}[t]
    \centering
    \includegraphics[width=\linewidth]{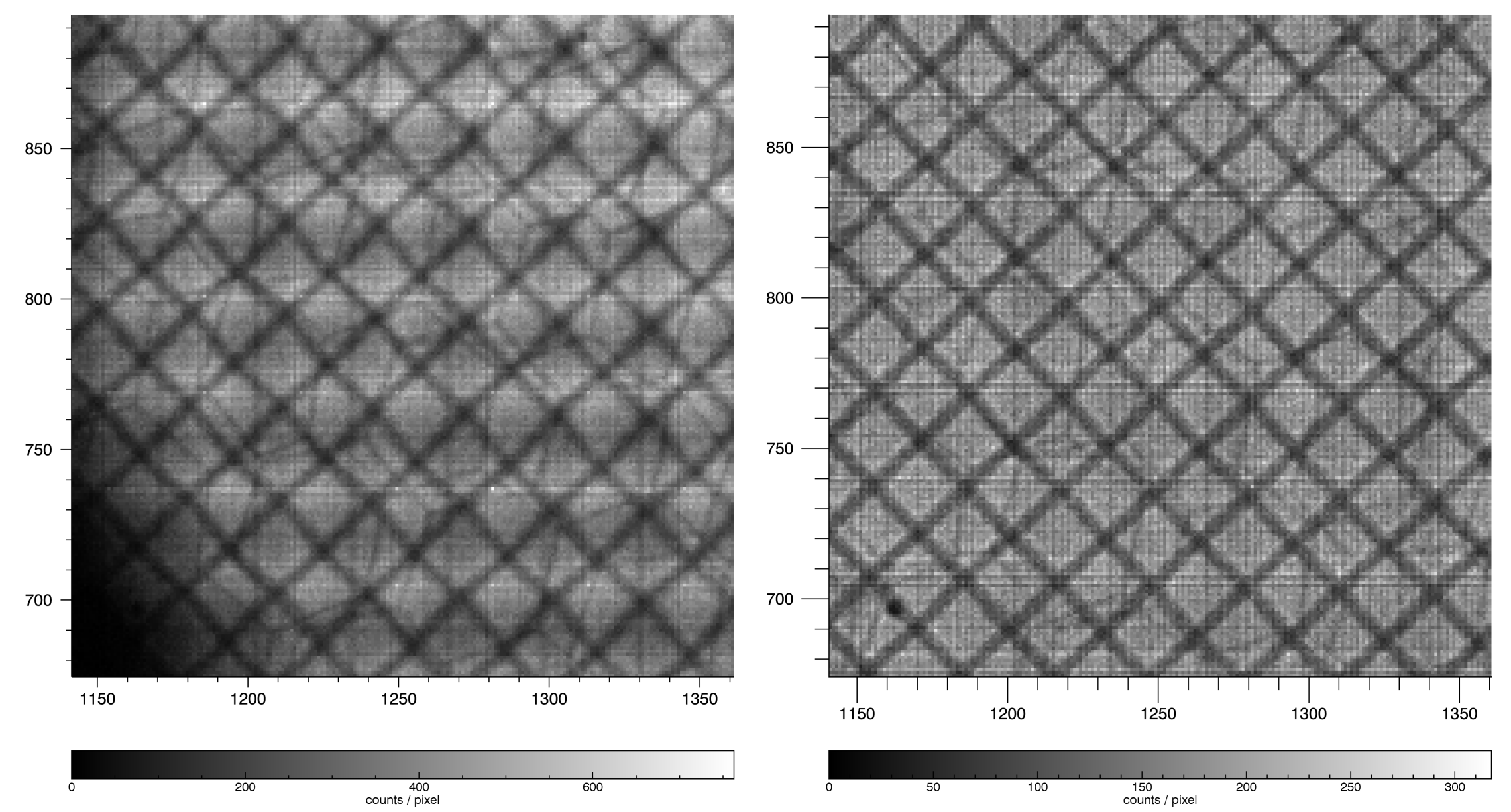}
    \caption{Comparison between the preliminary implementation of the noniterative centroiding in hardware, without (left) and with applied correction curve (right). Figure adapted from Diebold et al.\@ (2025)\cite{Diebold2025}.}
    \label{fig:corr_comp}
\end{figure}

In addition, systematic nonlinearities introduced by the centroid interpolation were investigated. These originate from the finite strip pitch and the fact that the measured charge distribution is not perfectly Gaussian. A correction curve was derived from flat-field measurements and implemented as a compact lookup table inside the FPGA. As presented in Figure~\ref{fig:corr_comp}, applying this correction significantly reduced periodic image artifacts and improved both geometric fidelity and image uniformity.

Overall, the work demonstrates that non-iterative Gaussian centroiding provides an attractive alternative to conventional Gaussian fitting. The method achieves nearly identical reconstruction quality while requiring substantially fewer computational resources, making it particularly well suited for high-throughput FPGA-based detector electronics. The full details of the implementation and all test results are published in Diebold et al.\@ (2025)\cite{Diebold2025}.

\section{CONCLUSIONS}
\label{sec:conc}

MCP-based detectors are a powerful option for UV instruments that provides unique features. Therefore, development and improvement of these detector systems are ongoing. The recent work at IAAT in this context focused on bringing opaque (Al)GaN photocathodes directly on MCPs and improving the quantum efficiency of semitransparent GaN on MgF$_2$ and sapphire.

Furthermore, for the readout of our co-planar cross-strip anode we implemented and tested an integer-based non-iterative centroiding algorithm; first in software, then on the FPGA of our readout electronics. Although the implementation in hardware almost reaches the performance of the algorithm in software - which in turn is significantly close to the benchmark of Gaussian fitting - optimization of the implementation is still ongoing.

\acknowledgments 
This work was supported by the Bundesministerium für Wirtschaft und Klimaschutz through the Deutsches Zentrum für Luft- und Raumfahrt e.V.\@ (DLR) under the grant number 50 QT 2001. Contacts to samples for ongoing studies of p-doping in Al$_{x}$Ga$_{1-x}$N films were prepared in an evaporation system funded partially by the Deutsche Forschungsgesellschaft (DFG INST 189/191-1 FUGG).

\bibliography{2026-SPIE-MCP} 

@Article{Caruana1986,
  author    = {Caruana, Richard A. and Searle, Roger B. and Heller, Thomas. and Shupack, Saul I.},
  journal   = {Analytical Chemistry},
  title     = {Fast algorithm for the resolution of spectra},
  year      = {1986},
  issn      = {1520-6882},
  month     = may,
  number    = {6},
  pages     = {1162--1167},
  volume    = {58},
  doi       = {10.1021/ac00297a041},
  publisher = {American Chemical Society (ACS)},
}

@Misc{Guo2012,
  author    = {Guo, Hongwei},
  month     = jun,
  title     = {A Simple Algorithm for Fitting a Gaussian Function},
  year      = {2012},
  doi       = {10.1002/9781118316948.ch31},
  isbn      = {9781118316948},
  journal   = {Streamlining Digital Signal Processing},
  pages     = {297--305},
  publisher = {Wiley},
}

@InBook{Diebold2022,
  author    = {Diebold, Sebastian},
  pages     = {1--36},
  publisher = {Springer Nature Singapore},
  title     = {{Proportional Counters and Microchannel Plates}},
  year      = {2022},
  isbn      = {9789811645440},
  month     = nov,
  booktitle = {Handbook of X-ray and Gamma-ray Astrophysics},
  doi       = {10.1007/978-981-16-4544-0_16-1},
}

@Article{Barstow2014,
  author    = {Barstow, M.A. and Casewell, S.L. and Holberg, J.B. and Kowalski, M.P.},
  journal   = {Advances in Space Research},
  title     = {{The status and future of EUV astronomy}},
  year      = {2014},
  issn      = {0273-1177},
  month     = mar,
  number    = {6},
  pages     = {1003--1013},
  volume    = {53},
  doi       = {10.1016/j.asr.2013.08.007},
  publisher = {Elsevier BV},
}

@InProceedings{Bougueroua2022,
  author       = {S. Bougueroua and M. {\AA}ngerman and J. Barnstedt and A. Colin and L. Conti and S. Diebold and R. Duffard and O. Janson and C. Kalkuhl and N. Kappelmann and T. Keilig and S. Klinkner and A. Krabbe and M. Lengowski and C. Lockowandt and P. Maier and T. M{\"u}ller and A. Pahler and T. Rauch and T. Schanz and B. Stelzer and M. Taheran and A. Vaerneus and K. Werner and J. Wolf},
  booktitle    = {Ground-based and Airborne Telescopes IX},
  title        = {{Status, flight preparation, and future instrument opportunities of the STUDIO balloon-borne telescope platform}},
  doi          = {10.1117/12.2628878},
  editor       = {Heather K. Marshall and Jason Spyromilio and Tomonori Usuda},
  organization = {International Society for Optics and Photonics},
  pages        = {121822N},
  publisher    = {SPIE},
  url          = {https://doi.org/10.1117/12.2628878},
  volume       = {12182},
  year         = {2022},
}

@InProceedings{Pahler2020,
  author       = {A. Pahler and M. {\AA}ngermann and J. Barnstedt and S. Bougueroua and A. Colin and L. Conti and S. Diebold and R. Duffard and M. Emberger and L. Hanke and C. Kalkuhl and N. Kappelmann and T. Keilig and S. Klinkner and A. Krabbe and O. Janson and M. Lengowski and C. Lockowandt and P. Maier and T. M{\"u}ller and T. Rauch and T. Schanz and B. Stelzer and M. Taheran and A. Vaerneus and K. Werner and J. Wolf},
  booktitle    = {Ground-based and Airborne Telescopes VIII},
  title        = {{Status of the STUDIO UV balloon mission and platform}},
  doi          = {10.1117/12.2575932},
  editor       = {Heather K. Marshall and Jason Spyromilio and Tomonori Usuda},
  organization = {International Society for Optics and Photonics},
  pages        = {114451Y},
  publisher    = {SPIE},
  url          = {https://doi.org/10.1117/12.2575932},
  volume       = {11445},
  year         = {2020},
}

@InProceedings{Vernoosfaderani2022,
  author       = {M. Taheran Vernoosfaderani and P. Maier and A. Pahler and S. Bougueroua and S. Klinkner and A. Krabbe and T. V{\"o}lker and J. Ackermann and M. {\AA}ngermann},
  booktitle    = {Software and Cyberinfrastructure for Astronomy VII},
  title        = {{Creating a highly flexible and autonomous stratospheric observatory: the essential elements of the European Stratospheric Balloon Observatory payload control software}},
  doi          = {10.1117/12.2629883},
  editor       = {Jorge Ibsen and Gianluca Chiozzi},
  organization = {International Society for Optics and Photonics},
  pages        = {1218908},
  publisher    = {SPIE},
  url          = {https://doi.org/10.1117/12.2629883},
  volume       = {12189},
  year         = {2022},
}

@InProceedings{Diebold2022a,
  author    = {Diebold, Sebastian J. and Barnstedt, Jürgen and Chandra, Bharat and Conti, Lauro and Ghatul, Shubham and Kappelmann, Norbert and Mohan, Rekhesh and Murthy, Jayant and Nair, Binukumar G. and Prabha, Shanti and Rai, Richa and Safonova, Margarita and Stelzer, Beate and Werner, Klaus},
  booktitle = {Space Telescopes and Instrumentation 2022: Ultraviolet to Gamma Ray},
  title     = {{TINI -- a mission for FUV spectroscopy of extended objects}},
  year      = {2022},
  editor    = {den Herder, Jan-Willem A. and Nakazawa, Kazuhiro and Nikzad, Shouleh},
  month     = aug,
  pages     = {112},
  publisher = {SPIE},
  doi       = {10.1117/12.2630121},
}

@Article{Muth1999,
  author    = {Muth, J. F. and Brown, J. D. and Johnson, M. A. L. and Yu, Zhonghai and Kolbas, R. M. and Cook, J. W. and Schetzina, J. F.},
  journal   = {MRS Internet Journal of Nitride Semiconductor Research},
  title     = {Absorption Coefficient and Refractive Index of GaN, AlN and AlGaN Alloys},
  year      = {1999},
  issn      = {1092-5783},
  number    = {S1},
  pages     = {502--507},
  volume    = {4},
  doi       = {10.1557/s1092578300002957},
  publisher = {Springer Science and Business Media LLC},
}

@Article{Curtis2025,
  author    = {Curtis, Travis and Tremsin, Anton and Siegmund, Oswald and McPhate, Jason and Nell, Nicholas and Tercero Macua, Dennis},
  journal   = {Journal of Astronomical Telescopes, Instruments, and Systems},
  title     = {MCP detectors: overview, advances, and prospects for Habitable Worlds Observatory},
  year      = {2025},
  issn      = {2329-4124},
  month     = jun,
  number    = {04},
  volume    = {11},
  doi       = {10.1117/1.jatis.11.4.042206},
  publisher = {SPIE-Intl Soc Optical Eng},
}

@article{Diebold2025,
  title={Progress towards a microchannel plate detector with {AlGaN} photocathode and cross-strip anode for ultraviolet astronomy},
  author={Diebold, Sebastian J and Barnstedt, Juergen and Conti, Lauro and Elsener, Hans-Rudolf and Hanke, Lars and H{\"o}ltzli, Markus and Kalkuhl, Christoph and Rau, Darleen J and Schaadt, Daniel M and Schanz, Thomas and Stelzer, Beate and Werner, Klaus},
  journal={Journal of Astronomical Telescopes, Instruments, and Systems},
  volume={11},
  number={4},
  pages={042230--042230},
  year={2025},
  publisher={Society of Photo-Optical Instrumentation Engineers}
}

@article{IBAD-MgO,
    author = {Wang, Siming and Antonakos, C. and Bordel, C. and Bouma, D. S. and Fischer, P. and Hellman, F.},
    title = {Ultrathin IBAD MgO films for epitaxial growth on amorphous substrates and sub-50 nm membranes},
    journal = {Applied Physics Letters},
    volume = {109},
    number = {19},
    pages = {191603},
    year = {2016},
    month = {11},
    issn = {0003-6951},
    doi = {10.1063/1.4966956},
    url = {https://doi.org/10.1063/1.4966956},
    eprint = {https://pubs.aip.org/aip/apl/article-pdf/doi/10.1063/1.4966956/14486417/191603_1_online.pdf},
}

@Article{Cai2021,
  author    = {Qing Cai and Haifan You and Hui Guo and Jin Wang and Bin Liu and Zili Xie and Dunjun Chen and Hai Lu and Youdou Zheng and Rong Zhang},
  journal   = {Light: Science + Applications},
  title     = {Progress On {AlGaN}-based Solar-blind Ultraviolet Photodetectors And Focal Plane Arrays},
  year      = {2021},
  month     = {12},
  number    = {1},
  volume    = {10},
  doi       = {10.1038/s41377-021-00527-4},
  publisher = {Springer Science and Business Media {LLC}},
}

@PhdThesis{Contiphd21,
  author    = {Conti, Lauro},
  title     = {{Development of UV MCP Detectors}},
  year      = {2021},
  type      = {phdthesis},
  copyright = {ubt-podok},
  doi       = {10.15496/PUBLIKATION-60938},
  language  = {en},
  publisher = {University of Tübingen},
  school={Universit{\"a}t T{\"u}bingen}
}

@InProceedings{Barstow2026,
  author    = {Martin A. Barstow AND Noora Alameri AND Costanza Argiroffi AND Miguel d'Avillez AND Nigel Bannister AND Joanna Barstow AND Rachel Bird AND Matthew Burleigh AND Sarah Casewell AND Elizabeth Clements AND Evangelina Deliporanidou AND Giulio Del Zanna AND Sebastian J. Diebold AND Jeremy J. Drake AND Vincent Fraux AND Michael Gillon AND Ana Ines Gómez de Castro AND Richard de Grijs AND Richard Hampson AND Graham Harper AND Aline Hermans AND Louise Harra AND Jérôme Jacobs AND Lionel Jacques AND Christian Kintziger AND Jon Lapington AND Domitilla de Martino AND Yaël Nazé AND Hamid Al-Naimiy AND Jonathan Nichols AND James Owen AND James Rogers AND Roisin Speight AND Beate Stelzer AND Venkatesh Sundararaman AND Mashhoor Al-Wardat AND Martin Townend AND Klaus Werner AND Allison Youngblood},
  booktitle = {Space Telescopes and Instrumentation 2026: Ultraviolet to Gamma Ray},
  title     = {{SIRIUS: stellar and ISM research via in-orbit ultraviolet spectroscopy}},
  year      = {2026},
  editor    = {den Herder, Jan-Willem A. and Nakazawa, Kazuhiro and Nikzad, Shouleh},
  publisher = {SPIE},
  doi       = {10.1117/12.3021533},
}

@article{Bowyer1994,
  author  = {Bowyer, Stuart and Lampton, Michael and Lewis, John and Wu, Xiaoyang and Jelinsky, Patrick and Malina, Roger F. and White, Nigel E. and others},
  title   = {The Extreme Ultraviolet Explorer Mission},
  journal = {Astrophysical Journal Supplement Series},
  volume  = {93},
  pages   = {569--583},
  year    = {1994}
}
\bibliographystyle{spiebib} 

\end{document}